# Title: Topological Modes in a Laser Cavity through Exceptional State Transfer

**Authors:** A. Schumer[1,2], Y. G. N. Liu[1], J. Leshin[3], L. Ding[1], Y. Alahmadi[3,4], A. U. Hassan[3], H. Nasari[1,3], S. Rotter[2], D. N. Christodoulides[3], P. LiKamWa[3], M. Khajavikhan[1]*

**Affiliations:**

[1]Ming Hsieh Department of Electrical and Computer Engineering, University of Southern California; Los Angeles, CA 90089, USA.

[2]Institute for Theoretical Physics, Vienna University of Technology (TU Wien); Vienna, A-1040, Austria.

[3]CREOL, The College of Optics and Photonics, University of Central Florida; Orlando, FL 32816-2700, USA.

[4]Center of Excellence for Telecomm Applications, Joint Centers of Excellence Program, King Abdul Aziz City for Science and Technology, Riyadh 11442, Saudi Arabia.

*Corresponding author. Email: khajavik@usc.edu

**Abstract:** Shaping the light emission characteristics of laser systems is of great importance in various areas of science and technology. In a typical lasing arrangement, the transverse spatial profile of a laser mode tends to remain self-similar throughout the entire cavity. Going beyond this paradigm, we demonstrate here how to shape a spatially evolving mode such that it faithfully settles into a pair of bi-orthogonal states at the two opposing facets of a laser cavity. This is achieved by purposely designing a structure that allows the lasing mode to encircle a non-Hermitian exceptional point while deliberately avoiding non-adiabatic jumps. The resulting state transfer reflects the unique topology of the associated Riemann surfaces associated with this singularity. Our approach provides a route to develop versatile mode selective active devices, and sheds light on the intriguing topological features of exceptional points.

**One-Sentence Summary:** A laser cavity is designed that operates in a topological mode whose spatial evolution exploits the topology surrounding an exceptional point.

**Main Text:**

The quantum adiabatic theorem, a corollary of the Schrödinger equation, provides remarkable insights into the behavior of slowly varying quantum systems. When the Hamiltonian gradually changes in time, the associated probability densities tend to evolve smoothly, thus allowing a quantum system to remain in its initial eigenstate. Interestingly, if this evolution follows a cyclic path around a spectral degeneracy, the related eigenvalue can acquire a geometric phase that solely depends on the traversed trajectory in parameter space (*1*, *2*). In condensed matter physics, when dealing with momentum space, it can be shown that the related concepts of Berry connection and curvature, that lift the path dependency of the observables, give rise to a host of fundamental topological properties like non-zero Chern number and integer quantum Hall conductance in solids (*3*).

Non-Hermitian systems and their spectral degeneracies, better known as exceptional points (EPs), have attracted attention in various physical disciplines, ranging from optics to electronics, optomechanics and acoustics (*4–14*). An intriguing feature of these non-Hermitian systems is that under appropriate conditions, their eigenvalues and corresponding eigenvectors tend to simultaneously coalesce, forming spectral degeneracies known as exceptional points (*4*, *15*). The presence of EPs not only affects a configuration that is statically operating in their vicinity, but it also alters the dynamical response of non-Hermitian systems. In contrast to a quasi-static encirclement of a Hermitian degeneracy (see Fig. 1C), cyclic parameter variations in non-Hermitian systems do not necessarily reproduce the input state (apart from a geometric phase) after completing a loop around an EP. Instead, a quasi-static cycle leads to a swap of the instantaneous eigenstates (see Fig. 1D) (*3*, *8*, *9*, *16*, *17*). Even more intriguing is the behavior of a non-Hermitian system when the EP-encirclement is carried out dynamically. In this latter case, the complex nature of the eigenvalues inhibits adiabatic evolution for all eigenvectors except for the one with the largest imaginary part of the corresponding eigenvalue due to non-adiabatic jumps (*17–19*).

Instead, these jumps produce a chiral behavior unique to non-Hermitian systems, in which the final state after a dynamic loop around an EP depends on the loop's winding direction rather than on the initial state at the loop's outset (see Fig. 1F) (*9, 19–25*). While this chiral behavior has recently been observed in a number of physical systems (*9, 20–22, 24, 26*), little has been done for exploiting this concept to establish a purely topological state in non-Hermitian configurations (*27–29*).

We introduce a type of topological mode appearing in non-Hermitian cavities that feature dynamical EP encirclement. In these systems, the interplay between the Riemann surfaces, the net gain, and gain saturation favors a spatially evolving lasing mode that morphs from one eigenstate profile to another while avoiding the aforementioned non-adiabatic jumps. As a result, we demonstrate a topologically operating single transverse mode laser, capable of simultaneously emitting in two different, but topologically linked transverse profiles, each from a different facet. Apart from its counterintuitive behavior, this laser constitutes an adiabatic non-Hermitian cavity that supports a fully topological resonant mode. The implementation of EP-encircling with gain, additionally avoids the considerable absorption losses that plagued previous reports of chiral state transfer with dissipative elements (*9, 20–22, 24, 26*). Furthermore, since the topological energy transfer relies solely on the adiabatic encircling of an EP degeneracy and not on the exact shape of the loop, the resulting lasing mode is robust against defects and fabrication imperfections, as well as fluctuations in gain [see (*30*) section 5 and 6].

Our laser cavity consists of two transversely coupled waveguides in a parity-time (PT) symmetric configuration, where one of the elements is subject to gain while the other one experiences loss (or a lower level of gain). A schematic of the device is shown in Fig. 2B alongside with SEM images in the insets of Fig. 2C. The dynamical encircling of the induced EP in time is implemented by modulating certain parameters of the structure along the propagation direction $z$. Specifically, by varying the coupling and the detuning between the two single mode waveguides in a continuous fashion, the system's transverse modes are steered around the EP as light circulates in the cavity. Each waveguide is accompanied by a neighboring strip that induces a change in its effective refractive index, providing the required detuning. These loading strips are intentionally designed not to be phase matched to the waveguide elements. The detuning between the two waveguides is thus determined by the distance between each waveguide and its adjacent strip and varies according to $s(z) = s_0 + (s_{min} - s_0)\sin(2\pi z/L)$ [see (30) section 1]. On the other hand, the dynamic coupling is attained by modulating the separation between the two primary waveguides, i.e., $d(z) = d_0 + (d_{max} - d_0)\sin(\pi z/L)$. Using the aforementioned modulation patterns, an EP-encircling loop is realized in parameter space when the light travels through the cavity once (half a cavity roundtrip), as shown in Fig. 2, A and B. The propagation direction of the wave through the cavity then determines the directionality of the EP-encircling. During a full cavity roundtrip, the EP is therefore encircled twice, once in each direction; the two loops of opposing directions in parameter space are chosen in such a way that non-adiabatic jumps are avoided (orange/purple line in left/right panel of Fig. 1F). It is this back-and-forth in the cavity that allows a single topological mode to be formed that is independent of the path taken in parameter space.

When a PT-symmetric pump profile is applied, in the absence of nonlinearities and saturation, the transverse mode evolution in the above active system is governed by a Schrödinger-type equation $i\partial_z \Psi(z) = H(z)\Psi(z)$ with $\Psi(z) = (E_1(z), E_2(z))^T$, where the z-dependent Hamiltonian is given by

$$H(z) = \begin{pmatrix} -\delta(z) - i\gamma + ig & \kappa(z) \\ \kappa(z) & \delta(z) - i\gamma \end{pmatrix}. \tag{1}$$

Here $\delta(z)$ is the detuning, $\kappa(z)$ is the coupling, $\gamma$ is the linear absorption loss, and $g$ is the gain provided through pumping. The instantaneous eigenvalues and eigenvectors of the Hamiltonian can be expressed as $\lambda_\pm(z) = i(g/2 - \gamma) \pm \kappa(z)\cos[\theta(z)]$ and $\Phi_\pm(z) = (2\cos[\theta(z)])^{-1/2} \left( e^{\pm i\theta(z)/2}, \pm e^{\mp i\theta(z)/2} \right)^T$, respectively, where $\theta(z) = \arcsin\left[\frac{g/2 + i\delta(z)}{\kappa(z)}\right] \in \mathbb{C}$. The PT-symmetry line is situated along $\delta = 0$ with the EP located at $\kappa_{EP} = g/2$, separating the PT-broken ($g/2 > \kappa$) from the PT-symmetric ($g/2 < \kappa$) phase. The start/end point of the EP-encircling section is at $\delta = 0$ and $\kappa \gg g/2$, and is chosen such that $\theta(0) = \theta_0 \approx g/2\kappa \ll 1$. Consequently, the eigenvector components are approximately equal in magnitude, which implies that the two supermodes emit with equal intensity in both waveguides at either facet. At $z = 0$ the relative phase $\varphi$ between the waveguide amplitudes of the supermodes are approximately $\varphi_-(0) \approx -\pi$ ($\pi$-out-of-phase) and $\varphi_+(0) \approx 0$ (in-phase) for $\Phi_-$ and $\Phi_+$, respectively. This is exactly reversed at the end of the encircling section ($z = L$), i.e., $\varphi_-(L) \approx 0$ and $\varphi_+(L) \approx -\pi$, such that the adiabatic following along the topological mode $\Phi_-(z)$ continuously morphs the transverse mode profile from being $\pi$-out-of-phase at one end to being in-phase at the opposite end of the cavity [for details see (*30*) section 5 and 6].

In order to intuitively understand the topological nature of this process, one may consider a random superposition of the transverse eigenvectors that are excited at one end of the encircling section of the device, after establishing the desired PT-symmetric pump profile. Irrespective of the initial excitation, by the end of a roundtrip in the cavity the state vector has undergone (at most) a single non-adiabatic transition towards the eigenvector subject to gain (purple/orange line in left/right panel of Fig. 1F) and is then caught in the adiabatic (fully topological) cycle, travelling back and forth between the facets; in fact, additional non-adiabatic transitions are forbidden as $\Phi_-$ is the amplified supermode throughout the entire length of the cavity. This transient behavior is simulated using the following nonlinear coupled stochastic differential equations, when excited through white noise $|\tilde{\eta}_1| \gg |\tilde{\eta}_2|$,

$$\frac{dE_1(\tilde{z})}{d\tilde{z}} = \frac{\tilde{g}_1 E_1(\tilde{z})}{1 + |E_1(\tilde{z})/E_s|^2} - \tilde{\gamma} E_1(\tilde{z}) + i\tilde{\delta}(\tilde{z})E_1(\tilde{z}) - i\tilde{\kappa}(\tilde{z})E_2(\tilde{z}) + \tilde{\eta}_1(\tilde{z}), \quad (2a)$$

$$\frac{dE_2(\tilde{z})}{d\tilde{z}} = -\tilde{\gamma} E_2(\tilde{z}) - i\tilde{\delta}(\tilde{z})E_2(\tilde{z}) - i\tilde{\kappa}(\tilde{z})E_1(\tilde{z}) + \tilde{\eta}_2(\tilde{z}), \quad (2b)$$

where $E_1(\tilde{z})$ and $E_2(\tilde{z})$ are the field amplitude in the waveguide subject to gain and loss, respectively, and $E_s$ is the saturation field. All the parameters are normalized with respect to the maximum coupling $\kappa_0 = \kappa(0) = \kappa(L)$, i.e., $\tilde{\gamma} = \gamma/\kappa_0$, $\tilde{\delta} = \delta/\kappa_0$, $\tilde{g}_1 = g_1/\kappa_0$, $\tilde{\kappa} = \kappa/\kappa_0$ and $\tilde{z} = \kappa_0 z$. After each passage through the cavity, the field amplitudes are reflected by the facets and travel through the system in the opposite direction. The back-and-forth propagation of 100 individual solutions to Eqs. 2a and 2b for a total of 6 cycles is shown in Fig. 3, A and B. We observe that any initial excitation is transferred towards the instantaneous eigenstate $\Phi_-$ within one cycle and the ensuing propagation follows this eigenvector adiabatically as the EP is repeatedly encircled in opposite direction. The relative phase between the waveguide amplitudes changes continuously from $-\pi$ to $0$ and back during a full roundtrip.

Finally, to obtain a self-consistent steady state lasing solution a Rigrod-type model was employed that considers the waves in both cavities travelling left-to-right and right-to-left simultaneously (*31*),

$$\frac{dE_1^\pm(\tilde{z})}{d\tilde{z}} = \pm\left(\frac{\tilde{g}_1 E_1^\pm(\tilde{z})}{1 + (|E_1^+(\tilde{z})/E_s|^2 + |E_1^-(\tilde{z})/E_s|^2)} - \tilde{\gamma}E_1^\pm(\tilde{z}) + i\tilde{\delta}(\tilde{z})E_1^\pm(\tilde{z}) - i\tilde{\kappa}(\tilde{z})E_2^\pm(\tilde{z})\right), \quad (3a)$$

$$\frac{dE_2^\pm(\tilde{z})}{d\tilde{z}} = \pm\left(-\tilde{\gamma}E_2^\pm(\tilde{z}) - i\tilde{\delta}(\tilde{z})E_2^\pm(\tilde{z}) - i\tilde{\kappa}(\tilde{z})E_1^\pm(\tilde{z})\right). \quad (3b)$$

Here the subscripts (1,2) refer to the first and second waveguide, and the superscripts correspond to the wave propagating from left-to-right $(+)$ and right-to-left $(-)$. The lasing modes have to replicate after each roundtrip within the resonator and obey the boundary conditions $E_i^+(0) = R_L E_i^-(0)$ and $E_i^-(\kappa_0 L) = R_R E_i^+(\kappa_0 L)$, where $R_L$ and $R_R$ are the reflectances at the left and right facet, respectively. After the transient behavior has settled in the instantaneous eigenvector $\Phi_-$, the dynamical encircling process is characterized solely by the topological adiabatic energy transfer between the two mode profiles at the output ports. The ratio of the field intensities in the two waveguides are equal at each facet (see Fig. 3D), while the relative phase of the state vector changes from $\varphi \approx -\pi$ at $z = 0$ to $\varphi \approx 0$ at $z = L$ (see Fig. 3C), corroborating that the system is lasing in the topological mode $\Phi_-$ [see also (*30*) section 9].

The laser structures are fabricated on an InP substrate wafer containing a 300 nm InGaAsP multiple quantum well active region that is covered with a 500 nm of epitaxially grown InP layer. The fabrication procedure for realizing the devices is described in (*30*) section 2. The structure comprises a 2 mm long encircling path after which the two loaded-strip waveguides are separated further to prevent additional coupling. In the main part, the width of each waveguide is 900 nm and the separation between the two waveguides varies between 600 nm to 1500 nm. The width of the detuning strips is 400 nm and their distance to the waveguides changes between 300 nm and 900 nm. The electromagnetic simulations of the modes, coupling strengths, and detunings can be found in (*30*) section 1. Because of the short free spectral range (FSR) of the cavities, 2 mm long grating mirrors based on sidewall modulation are incorporated at one end of the two waveguides to limit the number of longitudinal modes (Fig. 2, B and C). The gratings are identical (ridge widths: 1200 nm, period: 246 nm, and 50% duty cycle) and designed to promote spectrally narrow emission at ~1596 nm [see (*30*) section 3].

The fabricated laser structure is pumped with a 1064 nm pulsed beam, focused by a high magnification NIR objective and cylindrical lenses positioned before the sample. This produces a pump beam with a width of 8 $\mu$m and a length of 2 mm. By adjusting the position of the beam with respect to the pattern, one waveguide can be pumped with almost constant intensity over the entire length of the device, while the other is left with little to no pump energy. A PT-symmetric configuration is thus established, exhibiting an EP at the gain contrast value $g_1/2 = \kappa$. The level of gain contrast can be selected by changing the position of the pump beam. The in-plane emission from the edge facet of the laser is collected and imaged on a NIR camera and a spectrometer for further analysis. By changing the location of the waveguide facets with respect to the objective lens, one can maneuver between observing the near- and far-field intensity patterns in the camera. The details of the measurement station are described in (*30*) section 4.

In order to factor out the effect of the dissimilarities between the two ends of the structure, we alternately pump either the first or the second waveguide and collect the emission from the same facet. Changing the pump profile switches the order of CW and CCW encirclements in a roundtrip, thus enabling us to observe the cavity output from the two ends without requiring to switch the probed facet. After the encircling section, the two waveguides are gradually separated to a distance of 5 $\mu$m at the emitting end, thus allowing the observation of both near-field and far-field through our set-up configuration. Here, when the upper waveguide is pumped, the left-to-right propagating wave corresponds to dynamically encircling the EP in CW direction, leading to an in-phase emission profile at the designated facet, followed by a CCW winding that promotes the $\pi$-out-of-phase-mode on the other facet. This difference is particularly evident in the far-field intensity distribution, which shows a bright lobe in the center of the interference pattern for the in-phase mode (Fig. 4, A–C) when the first waveguide is pumped. By shifting the position of the pump light to the second waveguide, the EP-encircling direction is reversed, resulting in a situation equivalent to viewing the opposite facet. In this case, the $\pi$-out-of-phase supermode leads to a far-field pattern with a node at the center and two bright lobes around it (Fig. 4, D–F). In both cases, the near-field intensity patterns are similar (shown in Fig. 4, A and C), with the two waveguides emitting with nearly equal intensity (the slight difference is caused by the unequally-pumped separated regions). Together with the far-field profile, this confirms that the observed patterns belong to the desired in-phase and $\pi$-out-of-phase emission profiles of the corresponding topological mode [see also (*30*) section 10] . Finally, to verify that the structure is indeed lasing, the light-light curves are collected for both pump scenarios (Fig. 4G). The lasing spectra for both outputs are shown in Fig. 4H, with their peak wavelength occurring near 1596 nm. Unlike standard coupled waveguide lasers that tend to show frequency splitting, here the conversion from one state to the other along the cavity results in the same phase accumulation and resonance wavelength for both output states. Our device presents a demonstration of lasing through topological mode transfer. These lasing structures exhibit emission profiles that are robust to various parameter variations that tend to cause instabilities and temporal fluctuations in standard lasers. Extending this concept to larger arrays can result in laser systems with fast switching between various spatial supermodes by appropriately modulating the pump profile. Our work also provides a route to study topological effects in non-Hermitian systems by linking the elimination of non-adiabatic jumps to the formation of spatially evolving topological modes in laser cavities.

of short Fabry-Perot cavity formed by uniform fiber Bragg gratings. *Opt. Express.* **14**, 6394–6399 (2006).

**Acknowledgments**

**Funding:**

We gratefully acknowledge the financial support from DARPA (D18AP00058), Office of Naval Research (N00014-16-1-2640, N00014-18-1-2347, N00014-19-1-2052, N00014-20-1-2522, N00014-20-1-2789), Army Research Office (W911NF-17-1-0481), National Science Foundation (ECCS 1454531, DMR 1420620, ECCS 1757025, CBET 1805200, ECCS 2000538, ECCS 2011171), Air Force Office of Scientific Research (FA9550-14-1-0037, FA9550-20-1-0322, FA9550-21-1-0202), US–Israel Binational Science Foundation (BSF; 2016381), Jet Propulsion Laboratory (013385-00001), European Commission project NHQWAVE (MSCA-RISE 691209), and Austrian Science Fund (FWF) project WAVELAND (P32300).

**Author contributions:** The idea was conceptualized by MK, DC, SR, PL. The theoretical and experimental investigations were primarily done by AS, YL, JL, LD, YA, AH, HN. All authors contributed to the preparation of the manuscript.

**Competing interests:** Authors declare that they have no competing interests.

**Data and materials availability:** All data are available in the main text or the supplementary materials.

**Supplementary Materials**

Materials and Methods

Fig S1 – S13

References (32 – 36)

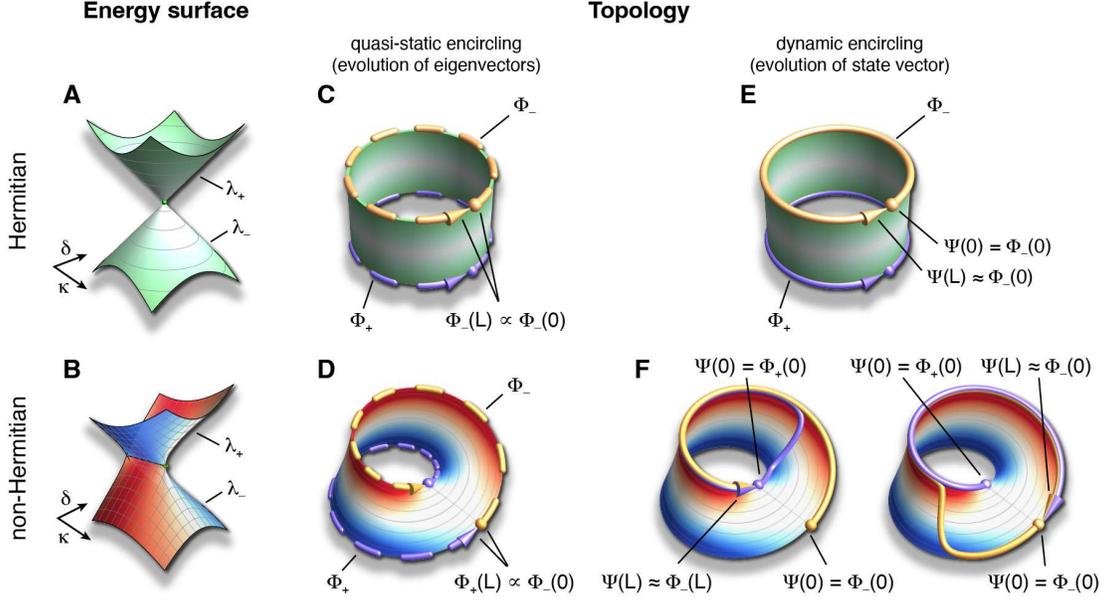

**Fig. 1. Encircling a Hermitian or a non-Hermitian degeneracy.** The occurrence of the interchange of the instantaneous eigenvectors when cycling around a degeneracy along a closed loop $C$ is independent of its shape, and only depends on the type of the enclosed degeneracy. (**A** and **B**) The energy surfaces of a Hermitian (top) and of a non-Hermitian degeneracy (bottom). The colors in (B, D and F) are connected to the imaginary part of the eigenvalues indicating gain ($\Im(\lambda) > 0$, red) and loss ($\Im(\lambda) < 0$, blue) behavior of the respective eigenvectors. (**C–F**) The topological equivalent of winding around a degeneracy. A cycle around a Hermitian degeneracy is represented by an untwisted closed sheet whereas a loop enclosing a non-Hermitian degeneracy corresponds to a Möbius strip. The eigenvector population $p(z) = (|c_+(z)|^2 - |c_-(z)|^2)/(|c_+(z)|^2 + |c_-(z)|^2)$, where $\Psi(z) = c_+(z)\Phi_+(z) + c_-(z)\Phi_-(z)$, is displayed on the vertical axis, such that the two instantaneous eigenstates $\Phi_\pm(z)$ lie on the edges of the sheets. (**C**) Quasi-statically winding around a Hermitian degeneracy along $C$ returns each eigenvector to itself. (**E**) Adiabatic cycling a Hermitian degeneracy starting from an eigenstate, e.g., $\Psi(0) = \Phi_-(0)$ (orange arrow), yields the same eigenvector after traversing $C$, i.e., $\Psi(L) \approx \Phi_-(0)$. (**D**)

Quasi-static evolution around an EP corresponds to the topology of a Möbius-strip as the eigenvectors interchange ($\Phi_{\pm}(0) \propto \Phi_{\mp}(L)$). (**F**) Upon dynamic EP-encircling in CW direction, any initial excitation (orange sphere: $\Psi(0) = \Phi_{-}(0)$; purple sphere: $\Psi(0) = \Phi_{+}(0)$) is transferred towards $\Phi_{-}$, such that after one cycle the state vector yields $\Psi(L) \approx \Phi_{-}(L) \propto \Phi_{+}(0)$ (left panel). When looping in CCW direction every initial state is again drawn to $\Phi_{-}$, but the state vector then gives $\Psi(L) \approx \Phi_{-}(0)$, leading to a chiral state transfer (right panel).

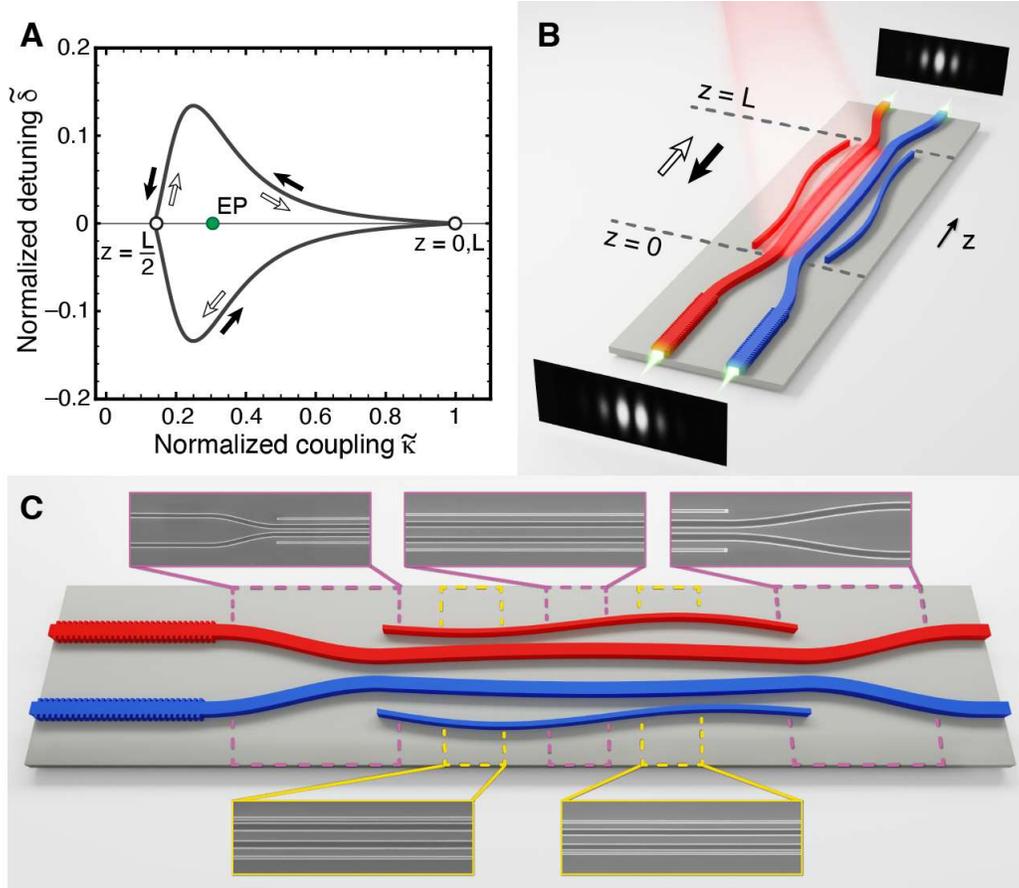

**Fig. 2. Operation principle and laser structure.** (**A**) Parameter path encircling the EP in the plane spanned by the normalized coupling $\tilde{k}$ and detuning $\tilde{\delta}$. (**B**) Illustration of the EP-encircling laser (not to scale). In addition to the losses due to absorption in both waveguides, the red waveguide experiences gain via optically pumping the encircling section of the cavity. The separation between the detuners and their respective main waveguides introduces detuning $\delta(z)$ while the separation between the two main waveguides generates coupling $\kappa(z)$. The grating reflector on the left end of each main waveguide acts as a wavelength filter. The steady state topological mode is characterized by the simultaneous emission of the in-phase (right end) and $\pi$-out-of-phase (left end) mode, each from one facet. (**C**) SEM images (small panels) of the laser

structure that demonstrate the variation of the separations between detuners as well as the main waveguides.

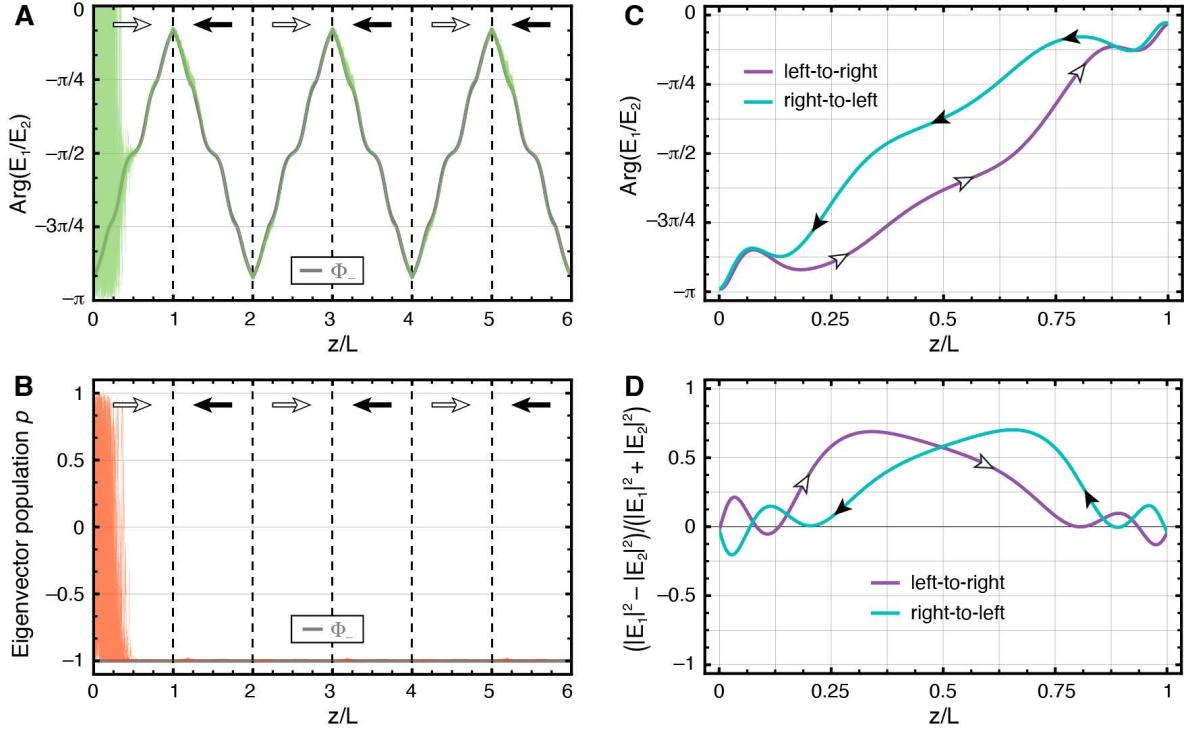

**Fig. 3. Numerically simulated transient and steady state behavior of the encircling part of the cavity.** (**A** and **B**) Numerical simulations of the transient field evolution for 6 passes in alternating directions through the cavity in the presence of gain saturation. In total 100 individual solutions of Eqs. 2a and 2b based on purely stochastic excitations are shown as thin green (A) and red (B) lines. The thick gray lines show the instantaneous eigenstates $\Phi_-(z)$ without noise. (A) The relative phase between the two waveguides evolves continuously from $-\pi$ to 0 and back within one cycle. (B) After an initial population transfer towards $\Phi_-$, the normalized eigenvector population $p$ shows that the ensuing adiabatic following of said eigenstate leads to the emission of different supermodes from each facet. (**C** and **D**) Evolution of the relative phase between the two cavities and the normalized intensity difference, respectively, of the left-to-right (purple) and right-to-left (cyan) travelling waves according to a Rigrod-type self-consistent simulation using Eqs. 3a and 3b showing the emission of different supermodes from each facet. The two spatial

supermodes are characterized by equal intensity in both waveguides and a phase difference that evolves from $-\pi$ to 0 and back.

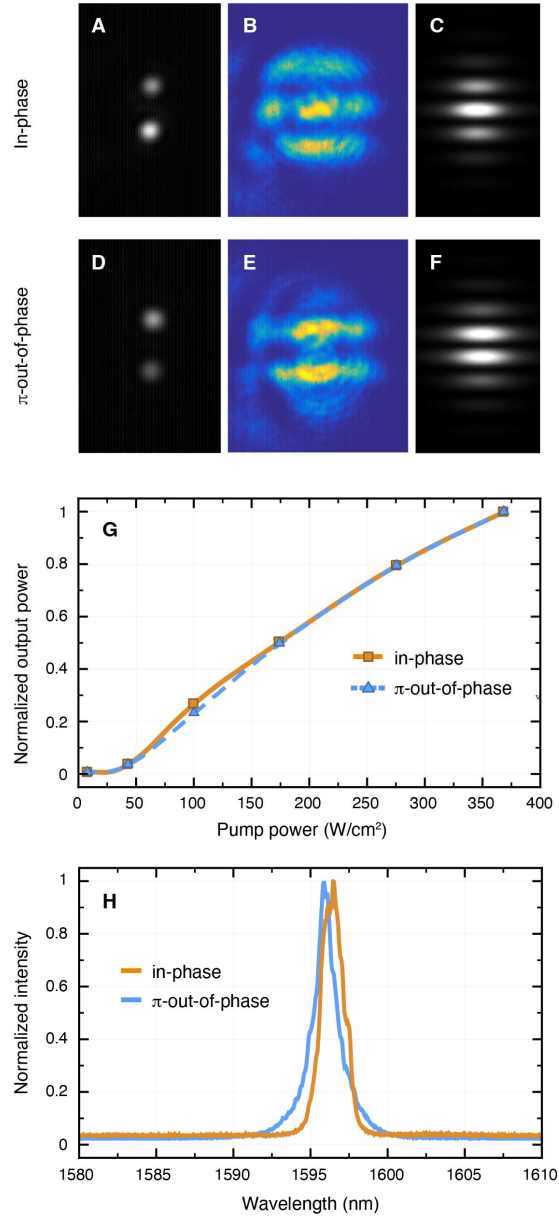

**Fig. 4. Near- and far-field intensity profiles, light-light curves and spectra.** (**A**, **B** and **C**) Experimental and simulation results, respectively, of the CW encircling scheme resulting in the in-phase supermode with a single bright central lobe. (**C**, **D** and **E**) Encircling the EP in CCW direction results in the emission of the π-out-of-phase supermode, which has a central dark spot between two bright lobes. The panels (A) and (D) show the respective near-field intensity profiles. Experimental far-field intensity distributions in (B) and (E) are colorized for a clearer visual characterization. Panels (C) and (F) show the simulated far-field intensity pattern. (**G**)

Normalized Light-light curves of the CW and CCW encirclement state, showing a characteristic lasing threshold. (**H**) Spectra of the CW and CCW encirclement setting.